\documentclass[twocolumn]{aastex63}
\usepackage{lineno}
\usepackage{epstopdf}
\usepackage{comment}
\usepackage{amsmath}
\shorttitle{A Stellar CME Candidate with High-confidence Level}
\shortauthors{Wang et al.}

\begin{document}

\title{Identifying a High-confidence Candidate of Stellar Coronal Mass Ejection on an M~dwarf}

\correspondingauthor{J. Wang}
\email{wj@nao.cas.cn}

\author{J. Wang} 
\affiliation{National Astronomical Observatories, Chinese Academy of Sciences, Beijing 100101,
China}

\author{J. R. Xu}
\affiliation{National Astronomical Observatories, Chinese Academy of Sciences, Beijing 100101,
China}

\author{Y. N. Ma}
\affiliation{School of Physics and Astronomy, Beijing Normal University, Beijing 100875, China}
\affiliation{National Astronomical Observatories, Chinese Academy of Sciences, Beijing 100101,
China}
\affiliation{School of Astronomy and Space Science, University of Chinese Academy of Sciences, Beijing, China}

\author{S. Jin}
\affiliation{Guangxi Key Laboratory for Relativistic Astrophysics, School of Physical Science and Technology, Guangxi University, Nanning 530004, China}
\affiliation{National Astronomical Observatories, Chinese Academy of Sciences, Beijing 100101,
China}
\affiliation{School of Astronomy and Space Science, University of Chinese Academy of Sciences, Beijing, China}

\author{L. J. Wang}
\affiliation{National Astronomical Observatories, Chinese Academy of Sciences, Beijing 100101,
China}

\author{Y. Xu}
\affiliation{National Astronomical Observatories, Chinese Academy of Sciences, Beijing 100101,
China}

\author{L. P. Xin}
\affiliation{National Astronomical Observatories, Chinese Academy of Sciences, Beijing 100101,
China}

\author{G. W. Li}
\affiliation{National Astronomical Observatories, Chinese Academy of Sciences, Beijing 100101,
China}

\author{Y. N. Zhu}
\affiliation{National Astronomical Observatories, Chinese Academy of Sciences, Beijing 100101,
China}

\author{H. Yang}
\affiliation{Institute de Recherche en Astrophysique \& Planétologie, 9 avenue du colonel Roche, 31028, Toulouse Cedex 04, France}

\author{X. H. Han}
\affiliation{National Astronomical Observatories, Chinese Academy of Sciences, Beijing 100101,
China}

\author{H. L. Li}
\affiliation{National Astronomical Observatories, Chinese Academy of Sciences, Beijing 100101,
China}

\author{Z. G. Dai}
\affiliation{Department of Astronomy, University of Science and Technology of China, Hefei 230026, China}
\affiliation{School of Astronomy and Space Science, Nanjing University, Nanjing, 210023, China}

\author{E. W. Liang}
\affiliation{Guangxi Key Laboratory for Relativistic Astrophysics, School of Physical Science and Technology, Guangxi University, Nanning 530004, China}

\author{X. Y. Wang}
\affiliation{School of Astronomy and Space Science, Nanjing University, Nanjing, 210023, China}

\author{C. Wu}
\affiliation{National Astronomical Observatories, Chinese Academy of Sciences, Beijing 100101,
China}

\author{J. Y. Wei}
\affiliation{National Astronomical Observatories, Chinese Academy of Sciences, Beijing 100101,
China}
\affiliation{School of Astronomy and Space Science, University of Chinese Academy of Sciences, Beijing, China}







\begin{abstract}

Although coronal mass ejections (CMEs) potentially play an important role in shaping the habitability of exoplanets, the 
firm detection of flaring-associated CMEs on remote stars remains a challenge task at current stage.
We here report time-resolved spectroscopic follow-up observations of a superflare (GWAC~251130A) on an M-dwarf, triggered by 
the SVOM/GWAC high-cadence transient survey. With an amplitude of $\approx2.5$~mag in $R-$band,
the bolometric flaring energy is estimated to be $1.2^{+1.0}_{-0.5}\times10^{35}$~erg.
The spectroscopy reveals complex flare-associated plasma motions.
A stellar CME candidate on the M-dwarf can be identified on the M~dwarf at high confidence, based on its maximum projected 
velocity ($V_{\mathrm{max}}\sim1500\ \mathrm{km\ s^{-1}}$) measured
from the blueshifted broad H$\alpha$ emission line-the second highest among all reported stellar CME candidates. This velocity makes
GWAC~251130A one of the only two events with a $V_\mathrm{max}$ exceeding twice
the surface escape velocity.
Both chromospheric evaporation and condensation can be additionally identified in the flare after the 
broad-line emission fades away. The CME mass and corresponding kinetic energy along the line-of-sight are estimated to be 
$\sim3.2\times10^{19}$~g and $\sim2.1\times10^{32}$~erg, respectively, 
which reinforces the slightly larger expelled mass with depressed kinetic energy in stellar CME than in the solar flare-CME events.
\end{abstract}



\section{Introduction} \label{sec:intro}

By analogy with the solar flares, stellar flares, releasing enormous energy of
$10^{28}-10^{39}$~erg over multiple wavelengths within a very short period of time, are commonly accepted 
to result from the magnetic reconnection process (see recent review by Kowalski 2024).
Instead of the $\alpha\Omega$ dynamo working well in the Sun, 
the $\alpha^2$ dynamo is believed to generate strong magnetic fields in cool main-sequence stars, 
owing to the absence of the boundary between the radiative and convective zones (e.g., Hotta et al. 2022; Bhatia et al. 2023).

The flares on M-dwarfs are of particular interest.
On the one hand, recent numerical simulations, together with laboratory experiments, 
suggest that the stellar flares
play an important role in shaping  the  habitability of exoplanets orbiting the host star  
(e.g., Kopparapu et al. 2013; Gao et al. 2026 and references therein).  
The intense UV photons and 
possible energetic material impulsively released in the flares are able to either tear off the atmosphere of an exoplanet or drive the production of greenhouse gas and hydrogen cyanide
(e.g.,  Cherenkov et al. 2017; Garcia-Sage et al. 2017; Airapetian et al. 2016, 2017, 2020; Tian et al. 2011; 
Chen et al. 2021; Rimmer et al. 2018; Tilley et al. 2019; Gao et al. 2026),
although the net effect remains an open question. On the other hand, the habitable zone of an 
M~dwarf is expected to
be quite close to the star, at a distance of only $\sim0.1$~au (Shields et al. 2016).

During a flare, a fraction of the confined plasma is expected to be expelled into interplanetary space by the erupting magnetic field lines,
resulting in the so-called coronal mass ejection (CME, e.g., Kahler 1992; Tsuneta 1996; Kliem et al. 2000; Karlicky \& Barta 2007;
Shibata \& Magara 2011; Li et al. 2016; Jiang et al. 2021), provided that the field eruption is strong enough and the overlying fields are
not too constraining (see the review by Forbes et al. 2006). There is indeed a linkage between flares and CMEs on 
the Sun (e.g., Yashiro et al. 2008; Aarnio et al. 2011; Webb \& Howard 2012).

Detecting a stellar CME nevertheless remains challenging with current
instruments largely because of their insufficient spatial resolution (e.g., reviews in Leitzinger \& Odert 2022; 
Moschou et al. 2019 and references therein). 
Even so, Houdebine et al. (1990) reported the fastest CME candidate to date 
on the active M~dwarf AD~Leo, 
in which a pronounced blue wing of the H$\gamma$ emission line yields a projected maximum velocity 
of $V_{\mathrm{max}}\sim5800\ \mathrm{km\ s^{-1}}$.  About a few dozen stellar CME candidates have
been reported on M~dwarfs by the Doppler effect of the ejected high-velocity material over past decade
(e.g., Argiroffi et al. 2019; Vida et al. 2019; Namekata et al. 2021, 2024; Koller et al. 2021;
Wang et al. 2021, 2022; Chen et al. 2022; Lu et al. 2022, 2025a,b;  Notsu et al. 2024; 
Cai et al. 2026; Cao \& Gu 2026), 
thanks to the great progress recently made in time-domain astronomy

In this paper, we identify a CME candidate at high-confidence, based on its maximum 
projected upward velocity as high as $\sim1500\ \mathrm{km\ s^{-1}}$ obtained from 
fast and time-resolved follow-up spectroscopy of a white light (WL) flare on an M~dwarf 
triggered by the 
SVOM/Ground-based Wide Angle Camera (GWAC) facilities
(Wei et al. 2016; Han et al. 2021; Cordier et al. 2026; Xin et al. 2026) that is 
designed not only to observe the optical prompt emission of $\gamma-$ray bursts (Xin et al. 2023),
but also to independently detect optical
transients with a high cadence.
Fast spectroscopy of about 10 flares triggered by the GWAC enabled us to identify four  
flare/CME candidates with a maximum projected velocity $\sim700-800\ \mathrm{km\ s^{-1}}$
during the 2021–2022 observational runs (Wang et al. 2021, 2022).

The paper is organized as follows. The discovery of the flare is described in Section 2. Section 3 presents the follow-up 
observations in both photometry and spectroscopy, along with the corresponding data reduction.  
The analysis and results are shown in Section 4. Section 5 gives the discussion and implications, with an emphasis on the stellar CME issue.

\section{Detection of the Transient by GWAC} \label{sec:transient}
 
After a period of upgrades and commissioning 
in both hardware and software from 2023 to the mid-2026, GWAC is now composed of 
10 mounts in total. Each 
mount is equipped with four Joint Field-of-View (JFoV) cameras and one Full Field-of-View (FFoV) camera. 
The field-of-view (FoV) and diameter are  $9\degr.8\times9\degr.8$ and 18~cm for each JFoV camera, respectively. 
The corresponding values are $64\degr\times42\degr$ and 3.5~cm for each FFoV camera. 
For the JFoV cameras, the limiting magnitude is 
determined to be 16th magnitude for a 10 seconds exposure in moonless phase.

A WL transient GWAC~251130A was triggered by the JFoV cameras
of GWAC at 2025-11-30T11:45:45 (UTC),
corresponding to $\mathrm{MJD= 61009.98773394}$ (hereafter $t_0$).
With the discovery images taken by GWAC, we can firmly rule out an artificial object,  
a small solar system body, and a known variable star. The transient, with a point-spread function
typical of nearby bright objects,
is located at a permanent celestial  position of $\mathrm{R.A. = 01^h32^m10^s.5}$ and 
$\mathrm{Decl = 36\degr13\arcmin38\arcsec}$ (J2000) with a typical uncertainty of 
$\sim2\arcsec$.
There are not only no known minor planets or comets\footnote{https://minorplanetcenter.net/cgi-bin/mpcheck.cgi?} 
brighter than $V = 20.0$~mag within a radius of 15\arcmin, 
but also no known variable stars or cataclysmic variables\footnote{https://vsx.aavso.org/index.php?view=search.top} around
the transient position within 1\arcmin.  

After the discovery of the transient, the raw images with an exposure time of 3~seconds taken by the JFoV cameras were then 
reduced by 
an offline pipeline. After the corrections of bias, dark, and flat-field, 
the pipeline first combined every 10 frames to enhance signal-to-noise ratio, and then performed
standard differential aperture photometry 
at the location of the transient and for several nearby bright reference
stars by using the \tt IRAF/APPHOT \rm package\footnote{IRAF is distributed by the National Optical Astronomical Observatories, which are operated by the Association of Universities for Research in Astronomy, Inc., under cooperative agreement with the National Science Foundation.}, in which an aperture size of 3 pixels was adopted in the photometry.
The absolute photometric calibration was finally performed against
the Gaia Data Release 3 catalog (Gaia Collaboration et al. 2022).


\section{Follow-up Observations, Data Reduction and Identification} \label{sec:ident}

\subsection{Photometric Follow-ups, Identification and Data Reduction}

After the trigger of GWAC~251130A, dedicated follow-up observations were carried out immediately
in photometry by the GWAC-F60A telescope with a diameter of 60~cm in the standard Johnson-Cousins $B-$ and 
$R-$bands. The monitoring was based on the dedicated real-time automatic transient validation
system (RAVS; Xu et al. 2020), which enables us not only to identify the
transient in minutes, but also to optimize sampling adaptively 
based upon the brightness and the evolution trend of the target. For the $B-$band, the exposure time is 
50 seconds for the first 12 frames, and then extended to 60 seconds for the remaining ones. 
The exposure time ranges from  20 to 60 seconds for the  $R-$band. 

An improved localization ($<1\arcsec$) given by the follow-up observations allows us to
identify the transient GWAC~251130A as a flare on the M~dwarf TIC~189437311 (=Gaia DR3 ~319048866740553728).
The basic properties of this host star are listed in Table~\ref{tab:properties}.
The values of $T_{\mathrm{eff}}$, $\log g$ and [Fe/H] listed in the table are estimated 
from the Gaia XP spectra by Huson et al. (2025). 
Owing to a lack of parallax measurement in the Gaia DR3 survey, we obtained  
the corresponding absolute $G-$band magnitude ($M_G$), stellar mass ($M_\star$) and radius ($R_\star$) for the star 
from its spectral type according to the modern mean stellar color and effective temperature sequence\footnote{https://github.com/dsavransky/MeanStars/blob/main/ MeanStars/EEM\_dwarf\_UBVIJHK\_colors\_Teff.txt} 
(Pecaut et al. 2012; Pecaut \& Mamajek 2013). Given the value of $M_G$, the distance is inferred
to be $\approx86$~pc from the measured 
apparent $G-$band brightness. Assuming spectral types of M3V and M4V instead yields
a distance range of 73$-$125\,pc for the object.

\begin{table}
        \centering
        \caption{Properties of the Host Starof GWAC~251130A (TIC~189437311)}
        \footnotesize
        \begin{tabular}{ccc} 
        \hline
        \hline
        Property & Value & Reference \\
          (1) & (2) & (3) \\     
        \hline 
       R.A. &  $\mathrm{01^h:32^m:10.^s9}$ & a\\
       Dec  &  +36\degr:13\arcmin:28\arcsec & a\\
       Sp. T & M3.5V & b\\
       $G-$band (mag) & $15.537\pm0.016$ & a\\
       GBp-GRp (mag) & $2.83$ & a \\
       $R-$band & 15.62 & c\\
       $V$ ($\mathrm{km\ s^{-1}}$) & $-7.58\pm4.90$ & a \\
       \hline
       $T_{\mathrm{eff}}$ (K) & $3289\pm8$ & d \\
       $\log(g/\mathrm{cm\ s^{-2}})$ & $4.83\pm0.01$ & d\\
       $\mathrm{[Fe/H]}$ &  $-0.03$ & d\\
       \hline
       Distance (pc) & $\approx86$ & e\\
       $M_G$ (mag) & 10.87 & e\\
       $M_{\star}$ ($M_\odot$)  &  $0.27$ & e \\
       $R_\star$ ($R_\odot$)& $0.30$ & e\\
        \hline
        \end{tabular}
        \tablecomments{References: a: Gaia Collaboration et al. (2022); b: Pecaut et al. 2012; 
        c: Monet et al. 2003; d: Huson et al. (2025); e: The values are estimated in this study. See the main text for the details.}
        \label{tab:properties}
\end{table}

\begin{figure}
      \centering
      \plotone{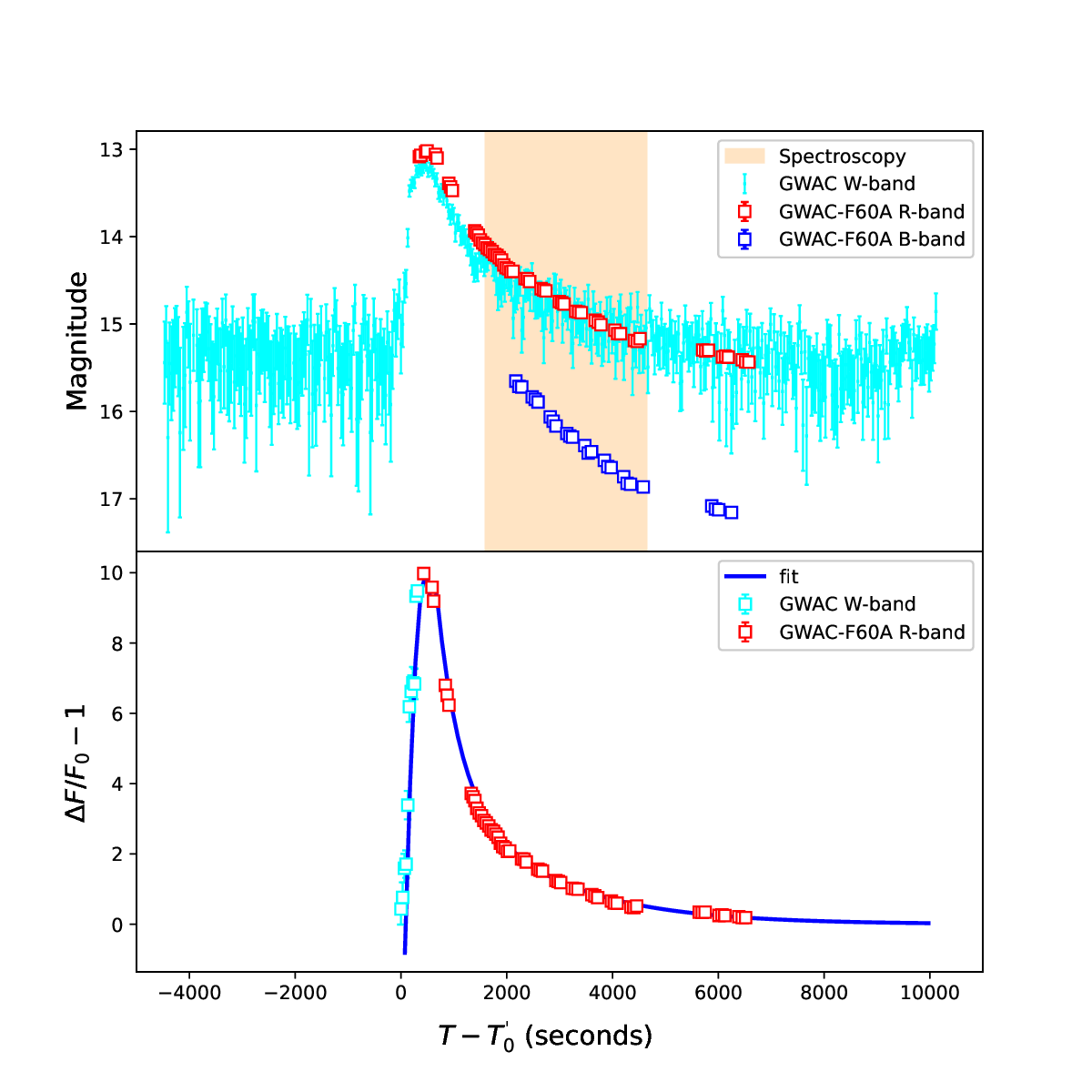}
      \caption{
   \it Upper panel: \rm: the light curves monitored by SVOM/GWAC and GWAC-F60A telescope for the stellar flare GWAC~251130A. 
   $T_0$ denotes the trigger time. 
   The duration of our monitor in spectroscopy is marked by the shadowed region in bisque.
   \it Lower panel:  \rm The best fit to the WL/$R-$band light curve of GWAC~251130A by a combination of a linear rise and 
   two exponential decay  components.}
\label{fig: lc}
\end{figure}

The raw images taken by the GWAC-F60A telescope were reduced by following the methods adopted 
for the GWAC observations (See Section 2 for the details). The upper panel of Figure \ref{fig: lc} shows the light curves 
of GWAC~251130A monitored by both GWAC and GWAC-F60A telescope. Two facts can be learned from the plot. 
First, the flare has been completely recorded by GWAC, covering both the fast rise and gradual 
decay phases. Secondly, there is a high consistency between the GWAC WL light curve and $R-$band 
light curve monitored by the GWAC-F60A telescope. Reddening can be safely neglected in the subsequent analysis becasue of its small color excess as low as $E(B-V) \approx 0.08$~mag. 
This color excess is estimated from the  
relationship $E(B-V)\approx 0.53\times(d/\mathrm{kpc})$ (Bohlin et al. 1978), 
assuming a hydrogen density around the Sun of $n_{\mathrm{H}} = 0.6\ \mathrm{atoms\ cm^{-3}}$ 
and a constant dust-to-gas ratio.

\subsection{Time-resolved Spectroscopy and Data Reduction}

After the discovery and quick identification of the transient, time-resolved long-slit spectroscopy was started with the
NAOC 2.16 m telescope (Fan et al. 2016) in the Target-of-Opportunity mode at $\sim28$~min after the trigger. 
A total of six spectra each with an exposure time of 600~seconds were obtained by 
the Beijing Faint Object Spectrograph and Camera equipped with a back-illuminated 
E2V55-30 AIMO CCD.  
The G8 grism was used in the spectroscopic observations, which provides a wavelength 
coverage from 5800 to 8200\AA. A slit width of 1\arcsec.8 oriented in the north-south 
direction finally resulted in a
spectral resolution of 3.5\AA\ as measured from the sky lines, which corresponds to a velocity of
$160\ \mathrm{km\ s^{-1}}$ for the H$\alpha$ emission line. The wavelength and flux 
calibrations were carried out with iron–argon comparison lamps and observations of Kitt Peak National Observatory standard 
star (Massey et al. 1988), respectively. All the spectra were taken quite close to
the meridian, with corresponding air mass ranging from 1.017 to 1.003. 
The duration of the time-resolved spectroscopy is marked by the region in bisque in the upper panel of Figure \ref{fig: lc}.

\begin{figure*}
      \centering
      \plotone{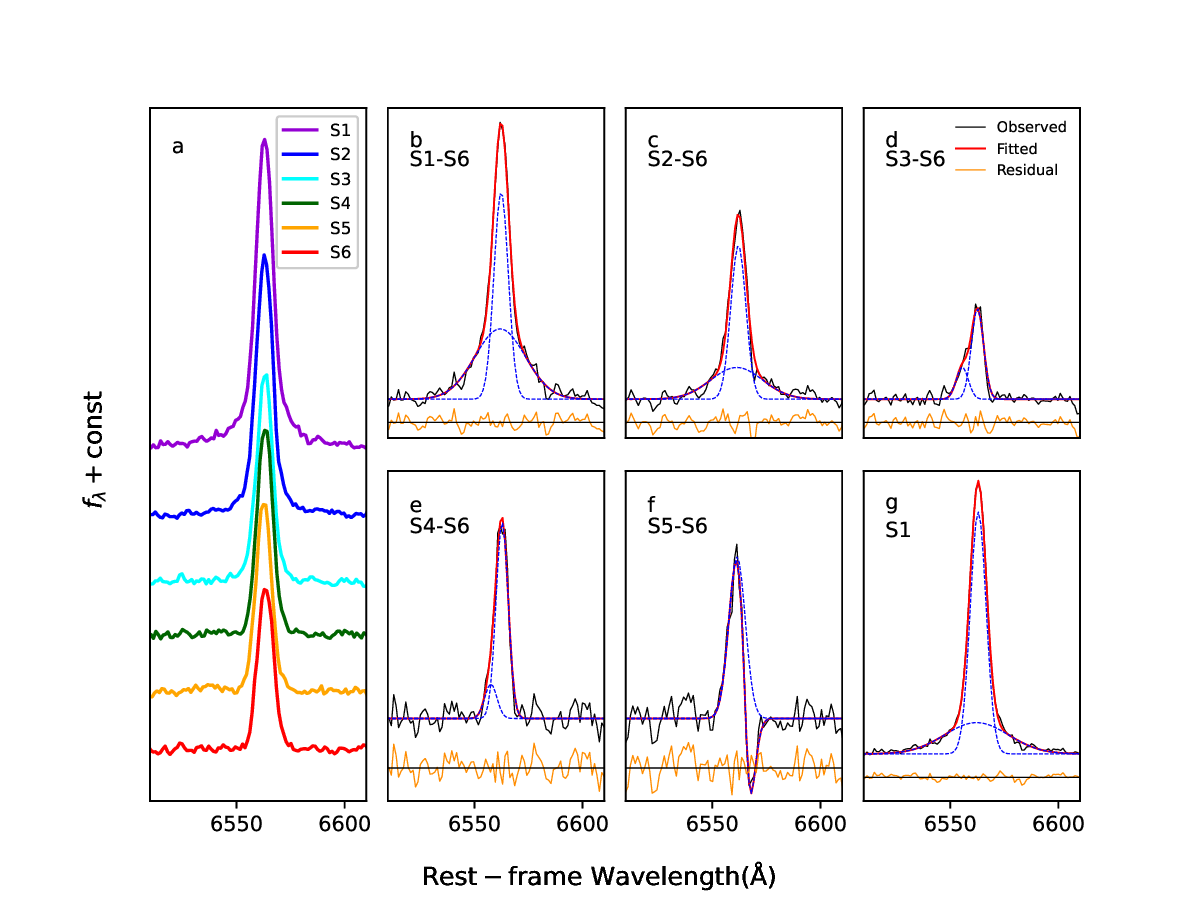}
      \caption{
   \it Panel (a): \rm Temporal spectral sequence of the H$\alpha$ emission-line sorted with time from top to bottom. 
   The spectra are shifted vertically by an arbitrary amount for clarity.
   \it Panels (b)-(f): \rm Modeling of differential H$\alpha$ line profiles by a linear combination of a set of Gaussian emission (or absorption)
   components. In each panel, a local underlying continuum has already been removed from the differential spectrum before the 
   line profile modeling.  The observed line profile is denoted by the black line, and the modeled one by the red solid lines. 
   Each Gaussian function is shown by a blue dashed line. Note that, in order to highlight the decayed H$\alpha$ emission line, the y-axis scale
   in the panels (e) and (f) is smaller than that in the panels (b)-(d). 
   \it Panel (g): \rm the same as Panels (b)-(f), but for the line profile without a subtraction 
   of the reference spectrum S6. 
   }
\label{fig: spec}
\end{figure*}

By using the \tt IRAF \rm package, we extracted the one-dimensional (1D) spectra from the raw images 
through standard procedures, including bias subtraction and flat-field correction.
In order to build differential spectra by direct subtraction (see Section 4.2 for details), 
all the 1D spectra, including both objects and standard stars, were extracted with fixed apertures of both object and sky emission. The extracted 1D spectra were then calibrated in wavelength and in flux by 
the corresponding comparison lamp and standard star, respectively.  
An alignment of the wavelength of sky [\ion{O}{1}]$\lambda$6300 emission line is adopted for 
a correction of the zero-point of the wavelength calibration, which leads to an accuracy of  wavelength calibration 
$\sim0.1$\AA, i.e., $\sim5\ \mathrm{km\ s^{-1}}$ at H$\alpha$ emission line. 
The spectral sequence is displayed in panel (a) of Figure \ref{fig: spec} for the H$\alpha$ region. 

\section{Analysis and Results}

\subsection{Optical Light Curve Modeling}

The lower panel of Figure \ref{fig: lc} shows the modeling of the optical light curve after combining the 
GWAC WL measurements at the early rise phase and the $R-$band measurements in the 
decaying phase after the peak. 
We adopt following widely used empirical model
(e.g., Davenport et al. 2014; Wang et al. 2022, 2026; 
Li et al. 2023; Xin et al. 2024): i.e., a linear function of  
 
\begin{equation}
  \Delta F = a_0+k_0t,\ t<t_{\mathrm{p}}
\end{equation}
for the early fast-rising phase and a sum of a set of exponential components

\begin{equation}
  \Delta F = \sum_{i=1}^{N} a_i\mathrm{e}^{-\frac{t}{\tau}},\  t\geq t_{\mathrm{p}}
\end{equation}
for the late gradual-decaying phase. $\Delta F = F/F_\mathrm{0} - 1$ is the relative flaring flux normalized to the 
quiescent level of $F_\mathrm{0} = (2.2\pm0.6)\times10^{-12}\ \mathrm{erg\ s^{-1}\ cm^{-2}}$ determined from its $R-$band brightness. The uncertainty is roughly evaluated from the 
canonical photometry accuracy of 0.34~mag reported in Monet et al. (2003). 
$t_\mathrm{p}$ is the corresponding peak time, and is simply fixed to the peak of GWAC
measurements because of its high cadence of 3~seconds.  

The best-fitting model described by a linear combination of a rising component and
two decaying components is overplotted in the lower panel of Figure \ref{fig: lc}. Based on the modeling, 
the equivalent duration (ED), 
defined as the time needed to emit the flare energy at the quiescent level, is calculated to be
$= 10200\pm100$~s by taking into account the uncertainty of the quiescent brightness. 
The fitting \rm results in a $R-$band energy of 
$E_R = 4\pi d^2\times ED\times F_{\mathrm{0}} = 2.0^{+1.7}_{-0.9}\times10^{34}$~erg, after 
proper error propagation.
By adopting a bolometric correction 
of $E_{\mathrm{bol}}/E_R\approx 6$ derived from a blackbody with a canonical temperature of $9000$~K 
(e.g., Kretzschmar 2011; Osten \& Wolk 2015; Corbett et al. 2025),
the total WL energy is inferred to be 
$E_{\mathrm{bol}}=1.2^{+1.0}_{-0.5}\times10^{35}$~erg, which 
means the event GWAC~251130A is typical of superflare.


\subsection{Emission-line Profile Modeling}

The panels (b)-(f)  of Figure \ref{fig: spec} show the modeling of the
differential H$\alpha$ emission-line profile at the first five epochs by the \tt IRAF/SPECFIT \rm task (Kriss 1994), in which the last spectrum is used as a reference in building the differential
spectra.   
In addition to a narrow Gaussian core, a strong broad Gaussian 
function is required to reproduce the differential H$\alpha$ line profiles at the two early epochs. 
At the following two epochs, an additional weak and blueshifted component is necessary for reproducing
the blue asymmetry of the differential profiles. 
A redshifted ``absorption'' is, however, needed to properly reproduce the differential profile at the last epoch.
As an additional test, we model the observed H$\alpha$ line profile at the first epoch 
by the same method, which confirms the strong broad H$\alpha$ emission with a line width of 
$\approx1500\ \mathrm{km\ s^{-1}}$ and is shown in the panel (g) of Figure \ref{fig: spec}.

The results are presented in Table~\ref{tab:profile} for the line profile modeling, in which all the uncertainties
correspond to the 1$\sigma$ significance level resulting from the modeling and a proper error propagation.
The reported line widths are corrected by the instrumental resolution through 
$\mathrm{FWHM}=\sqrt{\mathrm{FWHM}_{\mathrm{obs}}^2-\mathrm{FWHM}_{\mathrm{inst}}^2}$, where $\mathrm{FWHM}_{\mathrm{obs}}$
and $\mathrm{FWHM}_{\mathrm{inst}}$ are the observed line width and instrumental resolution, respectively.
$\Delta v$ is the bulk velocity shift of the additional Gaussian component  with respect to 
the central narrow core: $\Delta v = c\times(\lambda_{\mathrm{b}}-\lambda_{\mathrm{n}})/\lambda_{\mathrm{n}}$,
where $\lambda_{\mathrm{b}}$ and $\lambda_{\mathrm{n}}$ are the modeled line centers.   
According to this definition, a positive  and a negative $\Delta v$ denote a downflow and an upflow, respectively.

We further measure the maximum projected velocity $V_\mathrm{max}$ that is defined as the 
position where the H$\alpha$ high-velocity blue wing merges with the continuum. i.e., the line 
specific flux drops to the level of the 1$\sigma$ fluctuation of the continuum. 
From the observed H$\alpha$ line profile 
at the first epoch, $V_\mathrm{max}$ is measured to be 
$\approx1550\ \mathrm{km\ s^{-1}}$ in the S1 spectrum.
A slightly smaller value of $\approx1480\ \mathrm{km\ s^{-1}}$
can be alternatively obtained from a measurement in the S1$-$S6 differential spectrum. 
This discrepancy of the measured $V_\mathrm{max}$ is likely caused by an enhanced 
continuum fluctuations in the differential spectrum as expected for
a spectral subtraction.

In addition to H$\alpha$, an obvious flux decay can be identified for the \ion{He}{1}$\lambda$6678 emission line. 
In fact, the line is only detectable in the first differential spectrum, i.e., the S1$-$S6 spectrum. We measure the  \ion{He}{1}$\lambda$6678 
line flux in both S1 and  S1$-$S6 spectra by direct
integration with the \tt IRAF/SPLOT \rm task.
The flux, along with its uncertainty, is shown in Table~\ref{tab:profile}. The corresponding uncertainty
is calculated by $\sigma_{\mathrm{l}} = \sigma_{\mathrm{c}}\sqrt{N[1+EW/(N\Delta)]}$ (Perez-Montero \& Diaz 2013), 
where $\sigma_{\mathrm{c}}$ and $\Delta$ are the standard deviation of the continuum in a box near the line and 
the wavelength dispersion in units of \AA\,$\mathrm{pixel^{-1}}$, respectively. $EW$ is 
the equivalent width of the line, and $N$ the pixel number involved in the integration.

\begin{table}
\centering
        \caption{Results of H$\alpha$ Line-profile Modeling on the Differential spectra.}
        \label{tab:profile}
        \footnotesize
        \begin{tabular}{cccccc} 
         \hline
         \hline
          ID  & $f(\mathrm{H\alpha_n})$ & $f(\mathrm{H\alpha_b})$ &  $f(\mathrm{HeI\lambda6678})$  & $\mathrm{FWHM(H\alpha_b)}$ & $\Delta v$ \\
                 & \multicolumn{3}{c}{($\mathrm{10^{-15}\ erg\ s^{-1}\ cm^{-2}}$)} & \multicolumn{2}{c}{($\mathrm{km\ s^{-1}}$)} \\
          (1) & (2) & (3) & (4) & (5) & (6) \\
         \hline
          S1$^a$   & $236\pm4$ & $122.6\pm4.2$ & $14.8\pm0.7$ &  $1581\pm85$ & $-36\pm19$ \\
          S1$-$S6  & $89\pm4$ & $105.9\pm4.3$ &  $5.6\pm0.8$ & $1245\pm64$ & $-20\pm16$ \\       
          S2$-$S6 & $69\pm5$ & $49.0\pm4.3$ &  & $1278\pm201$ & $-36\pm44$ \\
          S3$-$S6  & $33\pm14$ & $11.5\pm4.5$ &  & $254\pm79$ & $-319\pm183$ \\
          S4$-$S6  & $34\pm9$ & $5.9\pm2.5$ &  & $256\pm89$ & $-238\pm196$ \\
          S5$-$S6  & $38\pm3$ & $13.6\pm0.4^b$ & & $217\pm29$ & $338\pm24$ \\
         \hline
         \end{tabular}
         \tablecomments{Column (1): the number of  the high-resolution spectra in the time series. 
         Columns (2) and (3):  the modeled H$\alpha$ emission-line fluxes of the central narrow core and additional broad/blueshifted component. Column (4): the line width of the additional broad/blueshifted component, after removing the instrumental profile. Column (5): relative velocity shift, see the main text for the definition. }
         \tablecomments{$^a$: Results based on the line-profile modeling or direct integration on the original spectrum.  $^b$: This value is 
         the  measured equivalent width in unit of \AA. }
\end{table}

\rm

\section{Discussion and Implication}

With an estimated bolometric energy of $1.2^{+1.0}_{-0.5}\times10^{35}$~erg, 
the WL transient GWAC~251130A can be identified as a superflare from a M-dwarf.
The monitor of the superflare in spectroscopy enables us to
obtain a maximum projected velocity as high as $\approx1500\ \mathrm{km\ s^{-1}}$ from 
the H$\alpha$ line emission, along with complex profile evolution.

\subsection{Physical Origin of the Observed H$\alpha$ Line Emission}

\subsubsection{CME/Prominence eruption}

We propose that the blueshifted broad H$\alpha$ emission observed at the first two epochs 
likely results from a 
flare-associated CME or prominence eruption. 
On the one hand, 
in the spectrum S1 (and also the S1$-$S6 differential spectrum), the measured value of $V_\mathrm{max}\approx1500\ \mathrm{km\ s^{-1}}$ is in fact 
much larger than the corresponding stellar surface escape velocity $v_{\mathrm{esc}}$. 
The relation of $v_{\mathrm{esc}}=618(M_\star/M_\odot)^{1/2}(R_\star/R_\odot)^{-1/2}\ \mathrm{km\ s^{-1}}$ 
predicts $v_{\mathrm{esc}}\approx 587\ \mathrm{km\ s^{-1}}$ for GWAC~251130A, given 
the stellar mass and radius tabulated in Table \ref{tab:properties}.

\begin{figure}
      \centering
      \plotone{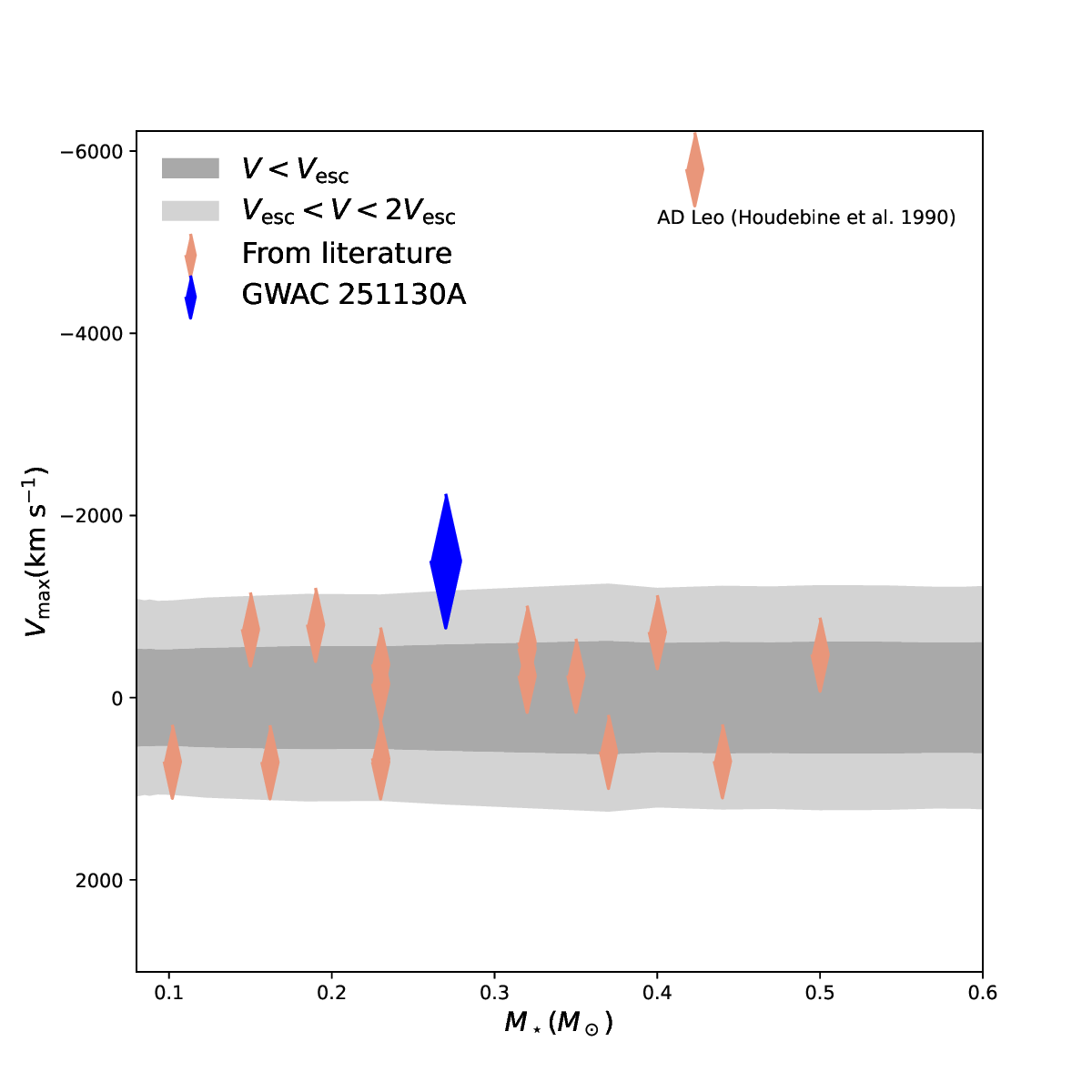}
      \caption{
      The maximum projected velocity ($V_{\mathrm{max}}$) measured from the high-velocity Balmer line wing plotted against the stellar mass for the CME candidates
      compiled from the literature (Houdebine et al. 1990; Wang et al. 2021, 2022; Koller et al. 2021; Lu et al. 2022, 2025a, b; Cai et al. 2026; Cao \& Gu 2026), along with the current case GWAC 251130A. The size of the symbols is arbitrary. 
      The dark and light gray regions mark the regions with 
      $V_{\mathrm{max}}<v_{\mathrm{esc}}$ and $v_{\mathrm{esc}}<V_{\mathrm{max}}<2v_{\mathrm{esc}}$, respectively, where 
      $v_{\mathrm{esc}}$ is the surface escape velocity calculated from the stellar mass and radius according to the  
      mean stellar color and effective temperature sequence given in Pecaut et al. (2012) and  Pecaut \& Mamajek (2013).}
\label{fig: vmax}
\end{figure}

Figure \ref{fig: vmax} compares the measured $V_\mathrm{max}$ between GWAC~251130A and 
other CME candidates compiled from literature. As shown in the figure, except for the extreme 
case on
AD Leo discovered in 1990 (Houdebine et al. 1990), GWAC~251130A is the only case with a measured $V_\mathrm{max}$ 
exceeding twice the corresponding surface escape velocity.

On the other hand, we argue that the Stark (pressure) effect is disfavored as an
explanation for
the blueshifted broad H$\alpha$ emission observed in the first two spectra.
The Stark (pressure) effect has long been 
proposed to explain the symmetrical broad Balmer emission detected at the early 
phase of some flares (e.g., Worden et al. 1984;
Hawley \& Pettersen 1991; Johns-Krull et al. 1997; Allred et al.
2006; Paulson et al. 2006; Kowalski et al.
2015, 2017; Namekata et al. 2020; Wu et al. 2022; Namizaki et al. 2023; 
Wang et al. 2026b).
Recent radiative hydrodynamic simulations show that the Stark broadening
depends on both flux and hardness of the injected high-energy 
non-thermal and thermal electron beams, which leads to 
an evolution of H$\alpha$ line width similar to that of WL flare during the 
impulsive and rapid-decay phases (e.g., Namekata et al. 2020).

However, there is a slight bulk blueshift for the broad H$\alpha$ line emission 
in GWAC~251130A (see the negative value of $\Delta v$ listed in Table \ref{tab:profile}). In addition,
the $V_\mathrm{max}$ measured in S1 spectrum corresponds to 
a H$\alpha$ broad-line width as high as $\approx66$\AA.
Based on the simulations in Namekata et al. (2020), 
an electron beam with a flux $\gg10^{12}\ \mathrm{erg\ s^{-1}\ cm^{-2}}$ and an extremely
hard energy spectrum is required to explain the broad H$\alpha$ emission 
in both S1 and S2 spectra by the Stark effect scenario. 
Therefore, based on the blueshift and large line width, the Stark effect is 
disfavored as an explanation for the broad H$\alpha$ emission.

\subsubsection{Chromospheric Evaporation and Condensation}

In the S3 and S4 spectra, the weak blueshifted H$\alpha$ emission with 
a bulk velocity $\approx200-300\ \mathrm{km\ s^{-1}}$ could result from a
chromospheric evaporation, whereas the redshifted absorption feature shown in 
panel (f) of Figure \ref{fig: spec} could be alternatively understood by
chromospheric condensation (or coronal rain), which results in 
an excess emission at the H$\alpha$ red wing in the S6 spectrum.

Both upward and downward chromospheric plasma motions driven by a flare have been indeed
reported for late-type main-sequence stars and RS CVn-type stars in previous studies (e.g., Honda et al. 2018; Koller et al. 2021; Wang et al. 2024, 2026; 
Cao \& Gu 2024, 2025).
When energetic electrons accelerated by the energy released in the magnetic reconnection are injected into chromosphere 
(e.g., Fisher et al. 1985; Canfield et al. 1990; Gunn et al. 1994; Innes et al. 1997; Fletcher et al. 2011; Berdyugina et al. 1999;  Tei et al. 2018; Li 2019;  
Tan et al. 2020; Chen et al. 2020; Yan et al. 2021), 
the chromospheric plasma is heated to a very high temperature. The heating results in 
an overpressure in the plasma. The overpressure drives the material either 
upward (e.g., Teriaca et al. 2003; Zhang et al. 2016; Brosius \& Daw 2015; 
Tian \& Chen 2018) or downward (e.g., Kamio
et al. 2005; Zhang et al. 2016; Libbrecht et al. 2019; Graham et al. 2020) motion.

The velocity of this motion depends on the heating rate of the electron beam.  
An ``explosive evaporation'' with a heating rate of $j_\mathrm{e}\geq3\times10^{10}\ \mathrm{erg\ s^{-1}\ cm^{-2}}$ 
is needed to generate the velocities of $\sim200-300\ \mathrm{km\ s^{-1}}$ revealed in GWAC~251130A (e.g., Milligan et al. 2006; Sadykov et al. 2015; Li 2019; Brosius \& Inglis 2017; Li et al. 2017).




\subsection{Mass and Kinetic Energy of Stellar CME}

We estimate the mass of stellar CME from the observed broad H$\alpha$ line emission following the widely accepted method 
(e.g., Houdebine et al. 1990; Wang et al. 2024, 2026; Cao \& Gu 2025; Wu et al. 2022): 

\begin{equation}
  M_{\mathrm{CME}}\geq N_{\mathrm{tot}}Vm_{\mathrm{H}}= \frac{L_{ji}m_{\mathrm{H}}}{A_{ji}h\nu_{ji}P_{\mathrm{esc}}}\frac{N_{\mathrm{tot}}}{N_j}
  \end{equation}
where $V$ is the total volume, assessed from the luminosity of a given emission line 
resulting from a transition from level $j$ to $i$
through $L_{ji}=N_jA_{ji}h\nu_{ji}VP_{\mathrm{esc}}$. $N_{\mathrm{tot}}$ and $N_j$ are the number densities of all hydrogen atoms and of the atoms at excitation level $j$, respectively.
$A_{ji}$ is the corresponding Einstein coefficient for a spontaneous decay, and $P_{\mathrm{esc}}$ the escape probability.

The H$\alpha$ line luminosity is first obtained from the flux of the   
modeled broad H$\alpha$ line emission in the S1 spectrum by $L_{\mathrm{H\alpha}}=4\pi d^2f_{\mathrm{H\alpha_b}}$, and then transformed to 
that of H$\gamma$ by assuming a Balmer decrement of three (Butler et al. 1988). 
Assuming $P_{\mathrm{esc}}=0.5$ then yields $M_{\mathrm{CME}} = 3.2^{+3.5}_{-0.9}\times10^{19}$~g, given
$A_{52}=2.53\times10^6\ \mathrm{s^{-1}}$ (Wiese \& Fuhr 2009) and $N_{\mathrm{tot}}/N_5= 2\times10^9$ according to 
the non-local thermal equilibrium modeling by Houdebine \& Doyle (1994a, b).
With the estimated CME  mass, the corresponding kinetic energy along the line-of-sight (LOS) axis is inferred to be 
$E_{\mathrm{k}}= \frac{1}{2}M_{\mathrm{CME}}(\Delta v)^2 = 2.1^{+8.0}_{-1.8}\times10^{32}$~erg, where $\Delta v= -36\pm19\  \mathrm{km\ s^{-1}}$ 
is the mean LOS velocity of the broad H$\alpha$ line emission measured in the S1 spectrum.
It is noted that the large uncertainties inferred for both $M_{\mathrm{CME}}$ and $E_{\mathrm{k}}$ are dominated by the distance uncertainty.

\begin{figure}
      \centering
      \plotone{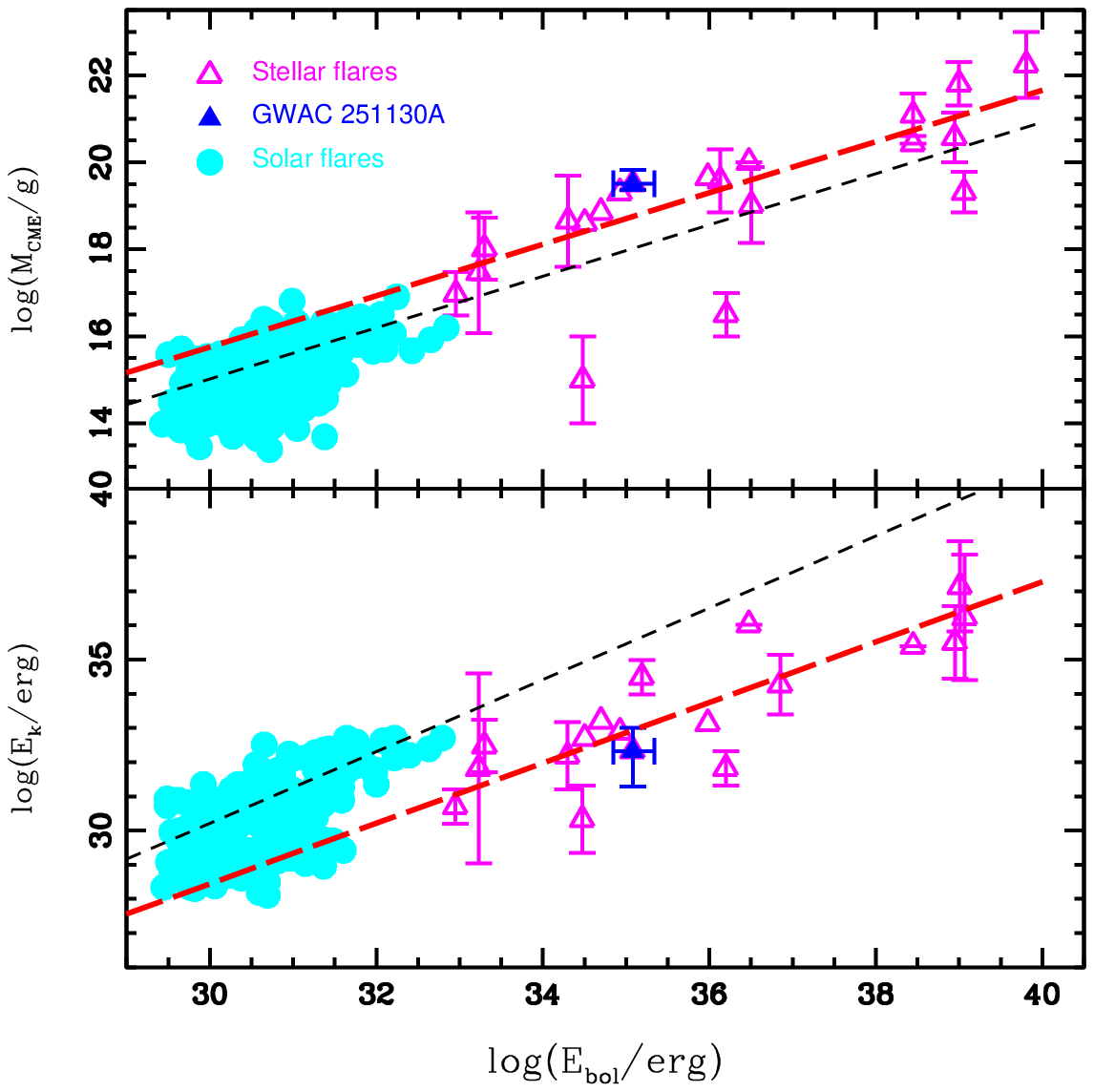}
      \caption{
   \it Upper panel: \rm mass of CME ($M_{\mathrm{CME}}$) plotted against the bolometric energy released in the flares 
   ($E_{\mathrm{bol}}$).  The magenta triangles 
   denote the stellar CME candidates including the GWAC~251130A reported in this study and the cases compiled from the literature 
   (Moschou et al. 2019; Argiroffi et al. 2019; Namekata et al. 2021; Wang et al. 2021, 2022; Wang 2023; Cao \& Gu 2026).  The solar flare-CME events studied in 
   Yashiro \& Gopalswamy (2009) are shown by cyan circles. The red  long-dashed line shows the best fit to 
   the stellar CME candidates (see Eq. 4 for the details), while the black short-dashed line the best fit to these solar events 
   (Drake et al. 2013).
   \it Lower panel: \rm the same as the upper one, but for the $E_{\mathrm{k}}$ (the kinetic energy along the LOS) 
   versus $E_{\mathrm{bol}}$ plot. }
\label{fig: enery}
\end{figure}

Figure \ref{fig: enery} shows  $E_{\mathrm{bol}}-M_{\mathrm{CME}}$  ( upper panel) and 
$E_{\mathrm{bol}}-E_{\mathrm{k}}$ (lower panel) plots for a sample of stellar CME
candidates and solar flare-CME events compiled from the literature, which yields the 
following best-fit relationships for the stellar CME candidates:

\begin{equation}
  \log\bigg(\frac{M_{\mathrm{CME}}}{\mathrm{g}}\bigg) =   (-2.0\pm0.8) + (0.59\pm0.12) 
  \log\bigg(\frac{E_{\mathrm{bol}}}{\mathrm{erg}}\bigg) 
\end{equation}
 and 
 
 \begin{equation}
  \log\bigg(\frac{E_{\mathrm{k}}}{\mathrm{erg}}\bigg) =  (1.9\pm0.7) + (0.88\pm0.12) 
  \log\bigg(\frac{E_{\mathrm{bol}}}{\mathrm{erg}}\bigg) 
\end{equation}
As shown in the plots, compared with the extrapolation of the detected solar CMEs,
the stellar CMEs tend to have slightly larger 
expelled masses with much lower kinetic energies. 
This kinetic-energy deficiency supports the proposed CME suppression mechanism
resulted from  a drag force done by a strong overlying magnetic field in late-type stars 
(e.g., Vrsnak et al. 2014; Zic et al. 2015;  Drake et al. 2016; Alvarado-Gomez et al. 2018).

%

\subsection{Flare Properties}

By assuming the flare duration is proportional 
to the Alfven travel time crossing the flare region, two scaling laws -- 
$\tau_\mathrm{d}\propto E^{1/3}B^{-5/3}$ and 
$\tau_\mathrm{d}\propto E^{-1/2}L^{5/2}$ -- have been derived 
by Maehara et al. (2015) and Namekata et al. (2017), where $\tau_\mathrm{d}$ is the 
flare duration defined as the $e-$folding decay time, $L$ and $B$ are 
the loop length and magnetic field, respectively. The modeled WL decay time
$\tau_{\mathrm{WL}}^{\mathrm{decay}} = 31.4\pm0.9$~min is used as a
proxy of $\tau_{\mathrm{d}}$, as is commonly done. 
With the bolometric
flaring energy released in GWAC~251130A, we have $L=5\times10^{10}\ \mathrm{cm} = 1.2R_\star$ and 
$B=140$~G.



\section{Conclusions}

The time-resolved spectroscopic follow-up of a WL superflare (GWAC~251130A) triggered by the SVOM/GWAC enables 
us to identify a candidate CME on an M~dwarf at high confidence, based on 
its second highest maximum projected velocity of $V_{\mathrm{max}}\sim1500\ \mathrm{km\ s^{-1}}$ measured from the blueshifted broad H$\alpha$ emission line. 
This high velocity makes GWAC~251130A one of the only two cases with measured $V_{\mathrm{max}}$ beyond twice the 
surface escape velocity.
Its large CME mass of $\sim3.2\times10^{19}$~g and corresponding small LOS kinetic energy 
of $\sim2.1\times10^{32}$~erg
reinforce the slightly larger expelled mass with depressed kinetic energy in stellar CME than in the solar flare-CMEs.
%
 \acknowledgments

We thank the anonymous referee for the helpful comments improving our study significantly.
This study is  supported by the National Key R\&D Program of China (grant Nos. 2024YFA1611604, 2024YFA1611702,
2024YFA1611700), by the Strategic Pioneer Program on Space Science, Chinese Academy of Sciences, grant
No. XDB0550401, and by  the Open Project Program of the Key Laboratory of Optical Astronomy,
NAOC, CAS.  We thank the night assistants and duty
astronomers of the NAOC 2.16 m telescope and the GWAC
system for their instrumental and observational help. Special
thanks go to the allocated observers for allowing us to interrupt
their observations at the 2.16 m telescope. This work has
made use of data from the European Space Agency (ESA)
mission Gaia (https://www.cosmos.esa. int/gaia), processed
by the Gaia Data Processing and Analysis Consortium (DPAC,
https://www.cosmos.esa.int/web/gaia/dpac/consortium). 
Funding for the DPAC has been provided by national institutions, in
particular the institutions participating in the Gaia Multilateral
Agreement. This research has made use of the VizieR catalog
access tool, CDS, Strasbourg, France (doi:10.26093/cds/vizier).
The original description of the VizieR service was published in
A\&AS 143, 23. 
\vspace{5mm}
\facilities{beijing: 2.16m}
\software{IRAF (Tody 1986, 1992), MATPLOTLIB}
          
\vskip12pt
\clearpage




\end{document}